\documentclass[prl,aps,twocolumn,showpacs,a4paper]{revtex4}

\usepackage{fullpage}
\usepackage{graphicx}

\begin{document}

\title{\bfseries Undulation instabilities in the meniscus of smectic membranes}
\author{J. C. Loudet$^1$, P. V. Dolganov$^2$, P. Patr\'{\i}cio$^{3,4}$, H. Saadaoui$^1$, and P. Cluzeau$^1$}

\affiliation{$^1$ Universit\'{e} Bordeaux 1, CNRS, Centre de Recherche Paul Pascal, Avenue A. Schweitzer F-33600
Pessac, France}

\affiliation{$^2$ Institute of Solid State Physics, Russian Academy
of Sciences, Moscow Region 142432 Chernogolovka, Russia}

\affiliation{$^3$ Centro de F\'{\i}sica Te\'{o}rica e Computacional,
Universidade de Lisboa, Avenida Professor Gama Pinto 2, P-1649-003
Lisboa Codex, Portugal}

\affiliation{$^4$ Instituto Superior de Engenharia de Lisboa, Rua
Conselheiro Em\'idio Navarro 1, P-1959-007 Lisboa, Portugal}

\date{\today}

\begin{abstract}

Using optical microscopy, phase shifting interferometry and atomic force microscopy, we demonstrate the
existence of undulated structures in the meniscus of ferroelectric smectic-C$^\ast$ films. The meniscus is
characterized by a periodic undulation of the smectic-air interface, which manifests itself in a striped
pattern. The instability disappears in the untilted smectic-A phase. The modulation amplitude and wavelength
both depend on meniscus thickness. We study the temperature evolution of the structure and propose a simple
model that accounts for the observed undulations.

\end{abstract}


\maketitle

When in contact to a solid substrate, isotropic liquids form a meniscus whose properties are controlled by the
well-established laws of capillarity and gravity \cite{rowlinson}. Wetting of liquids on surfaces, for example,
is of ubiquitous importance in many industrial fields and our everyday life experience. The case of complex
liquids, such as liquid crystals (LC), is much more complicated because of the partial ordering of molecules.
The molecular order breaks the rotationnal and/or translationnal symmetries of ordinary disordered liquids and
imparts elastic properties. As a result, most physical properties become anisotropic. Studies of wetting of
liquid crystals date back to the 70's, and the more specific case of smectic free-standing films (FSF), or
smectic membranes, was addressed more recently \cite{pieranski93,picano00,poniewierski02}. In contrast to
isotropic liquids, where the meniscus profile is exponential (e.g. soap films), the meniscus in smectic-A (smA)
FSF has a circular profile and forms a finite angle with the film (Fig. 1). The meniscus shape determines the
film disjoining pressure \cite{oswpier}, which governs its stability and is essential in many interesting
phenomena, such as thinning transitions \cite{stoebe94}. The shape and structure of menisci also play a key role
in capillary interactions between small solid objects trapped at fluid interfaces
\cite{kral00,loudet05,oettel08}. When embedded in FSF, inclusions self-organize due to both film deformations
and elastic distortions of the LC host \cite{poulin97,lubensky98,dolganov08,bohley08}.

The situation is even more complex in smectic-C (smC) or chiral smectic-C* (smC$^\ast$) phases
\cite{oswpier,degennes} where the molecules are tilted with respect to the layer normal. Projection of the
molecular axes onto the layer plane, defined as the $\mathbf{c}$-director, can vary in orientation and form
modulated structures. Meyer and Pershan \cite{meyer73} described the so-called splay domains in the meniscus of
smC$^\ast$ films, which were associated with surface-induced spontaneous polarization resulting in
$\mathbf{c}$-director splay. Layer undulation was proposed as a reason for the observed patterns around solid
inclusions in smA films \cite{conradi06}. Quite recently, Harth and Stannarius \cite{harth09} reported studies
of modulated striped structures in the menisci at the film edge as well as around inclusions embedded in films.
They interpreted the observed structures to $\mathbf{c}$-director distortions, consistent with splay domains.
Nevertheless, in spite of a long history of studies \cite{meyer73,conradi06,harth09}, uncertainty still exists
about the nature of these modulated structures.

\begin{center}
\begin{figure}[h]
  \begin{center}
    \includegraphics[width=2.6in]{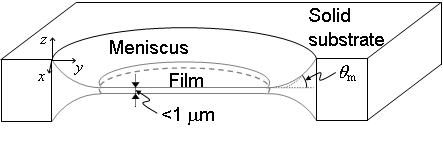}
  \end{center}
  \caption{Geometry of freely suspended smectic films.}\label{1}
\end{figure}
\end{center}

In this paper we demonstrate that stripes observed in smC$^\ast$ meniscus are related to undulations of the
smectic-air interface. The phenomenon resembles the formation of wrinkles at interfaces observed in various soft
matter systems \cite{huang07,vandeparre10,holmes10}, which attracts considerable attention. The observed
modulation crucially depends on temperature and meniscus thickness. A simple model, based on dilation-induced
strain applied to the smectic layers, is shown to be consistent with the experimental results.

We used a variety of experimental methods, namely polarized light microscopy, phase shifting interferometry
(PSI) and atomic force microscopy (AFM). Experiments were performed on several compounds, including
ferroelectric materials ZLI-3488 \cite{lalanne91}, SCE-9 and SCE-12 (Merck, England). Films in the smC$^\ast$
and smA phases were prepared on a hole in a thin glass plate (Fig. 1). After preparation, the films were kept at
a constant temperature for several hours to ensure the relaxation of the meniscus shape. Temperature was
controlled by a Mettler heating stage.

Figure 2 shows the characteristic patterns in the film meniscus. Nearly vertical lines are the boundaries
between the film and the meniscus (respectively, the left and right part in each figure). The meniscus thickness
increases from left to right. In the SmC$^\ast$ phase, various modulated structures develop only in the meniscus
and not in the film. First, a system of branched and then parallel stripes is formed [Fig. 2(a-c)]. We will
denote this modulated structure as a 1D pattern. In the thick part of the meniscus, a 2D pattern is observed. It
may be described as rows of square domains that appear for a well-defined meniscus thickness \cite{2dnet}. A
crossover region exists between the 1D and 2D patterns [Fig. 2(b)]: while the stripes are still visible, a
secondary modulation develops along the film thickness gradient. The period of the stripes and the size of the
squares increase with meniscus thickness. As already thoroughly described in \cite{harth09}, the meniscus
texture is independent of the chiral nature of the considered compounds but varies strongly with temperature. We
remind here that, in free-standing films, the smC$^\ast$ structure can exist above the bulk smC$^\ast$-smA
transition temperature, $T_C$, due to surface ordering \cite{oswpier}. On heating close to $T_C$ (Fig. 2), the
contrast of the stripe pattern decreases gradually. Just above $T_C$, residual stripes remain in the meniscus
[Fig. 2(c)]. They eventually disappear completely upon further heating while the square pattern partially
remains [Fig. 2(d)]. The 2D pattern fades away in a stepwise manner: the squares disappear row by row, following
an unzipping-like mechanism, starting from the thinner regions. Well above $T_C$, square domains are no longer
present and the whole meniscus becomes defect-free.

On cooling back to the smC$^\ast$ phase, the same patterns reappear in reverse order. The observed phenomena are
fully reproducible over heating/cooling cycles. At constant temperature, all the above structures survive for
many days as long as the film is stable.

\begin{center}
\begin{figure}[t]
  \begin{center}
    \includegraphics[width=3.0in]{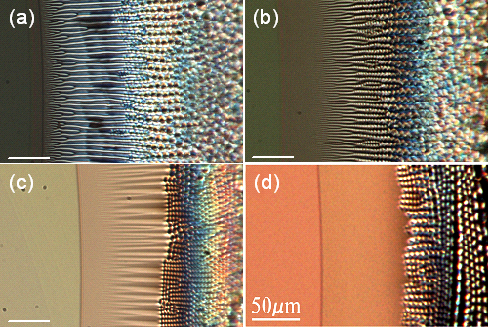}
  \end{center}
  \caption{Optical photographs showing the evolution of the meniscus
  structure with temperature (SCE-9 and SCE-12 compounds). (a)
  smC$^\ast$ phase, $T=37.5^\circ$C. (b) $T=67^\circ$C, still in the
  smC$^\ast$ phase. (c) $T=70^\circ$C, just above the bulk
  smC$^\ast$-smA phase transition. (d) $T=74.5^\circ$C, deep in the
  smA phase. All the photographs were taken in transmitted light
  with decrossed polarizers and various decrossing angles to enhance
  contrast.}\label{2}
\end{figure}
\end{center}

To clarify the nature of the observed structures, we used Phase Shifting Interferometry (PSI) to probe the
smectic-air meniscus profile, $z=u(x,y)$, with a resolution of a few nanometers. The principles of PSI and the
experimental setup were described elsewhere \cite{loudet06,harasaki,robinson}.

\begin{center}
\begin{figure}[t]
  \begin{center}
    \includegraphics[width=2.5in]{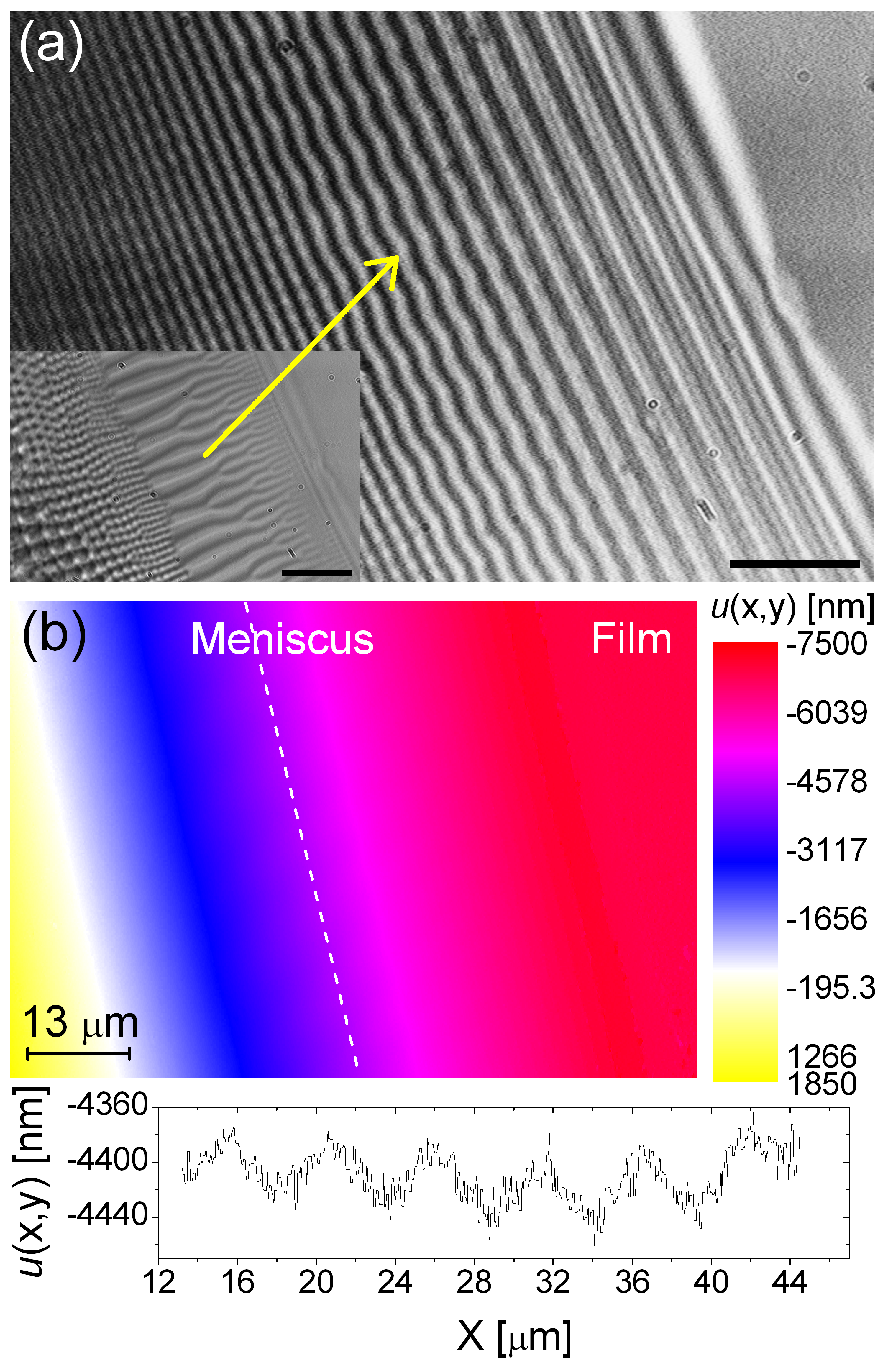}
  \end{center}
  \caption{(a) Interferogram of a ZLI-3488 smC$^\ast$ film
  (reflection mode). $T=40^\circ$C, scale bar: $13\,\mu$m. Insert:
  Corresponding optical photograph in transmitted white light. Scale
  bar: $16.2\,\mu$m. The arrow indicates the location of broad
  wrinkles in the interferogram. (b) Associated color-coded image
  plot of $u(x,y)$. The bottom graph shows the undulated meniscus
  profile along the oblique dashed line (region of broad
  stripes).}\label{3}
\end{figure}
\end{center}

Figure 3(a) shows an interferogram of a smC$^\ast$ meniscus. The main feature of this image is that interference
fringes are not straight but wavy in the stripes area. Therefore, the meniscus is not smooth (as in smA) but
rather undulated in this area. The corresponding height profile $u(x,y)$, reconstructed from a series of
interferograms, is presented in Fig. 3(b). The meniscus rises about $\simeq8\,\mu$m above the flat film. The
oblique dashed line marks the location of the broadest wrinkles in the inset of Fig. 3(a) while the bottom graph
exhibits the associated profile. In this region, undulations occur with amplitude around $\sim40-50\,$nm and
wavelength $\lambda\sim5\,\mu$m. The existence of wave-like fringes is directly correlated to the presence of
stripes: on heating above $T_C$, as the stripes disappear, the fringes smoothen and become straight. Close to
the flat film region, the fringes turn to wavelets with smaller amplitude and wavelength. In this area, the
amplitude of the wrinkles cannot be resolved as it falls within our experimental accuracy ($\sim5\,$nm). In the
thick part of the meniscus, preliminary PSI analyses of the 2D square lattice reveal that the interface
undulates in two orthogonal directions. A detailed analysis of these data will be given elsewhere.

\begin{center}
\begin{figure}[b]
  \begin{center}
    \includegraphics[width=2.5in]{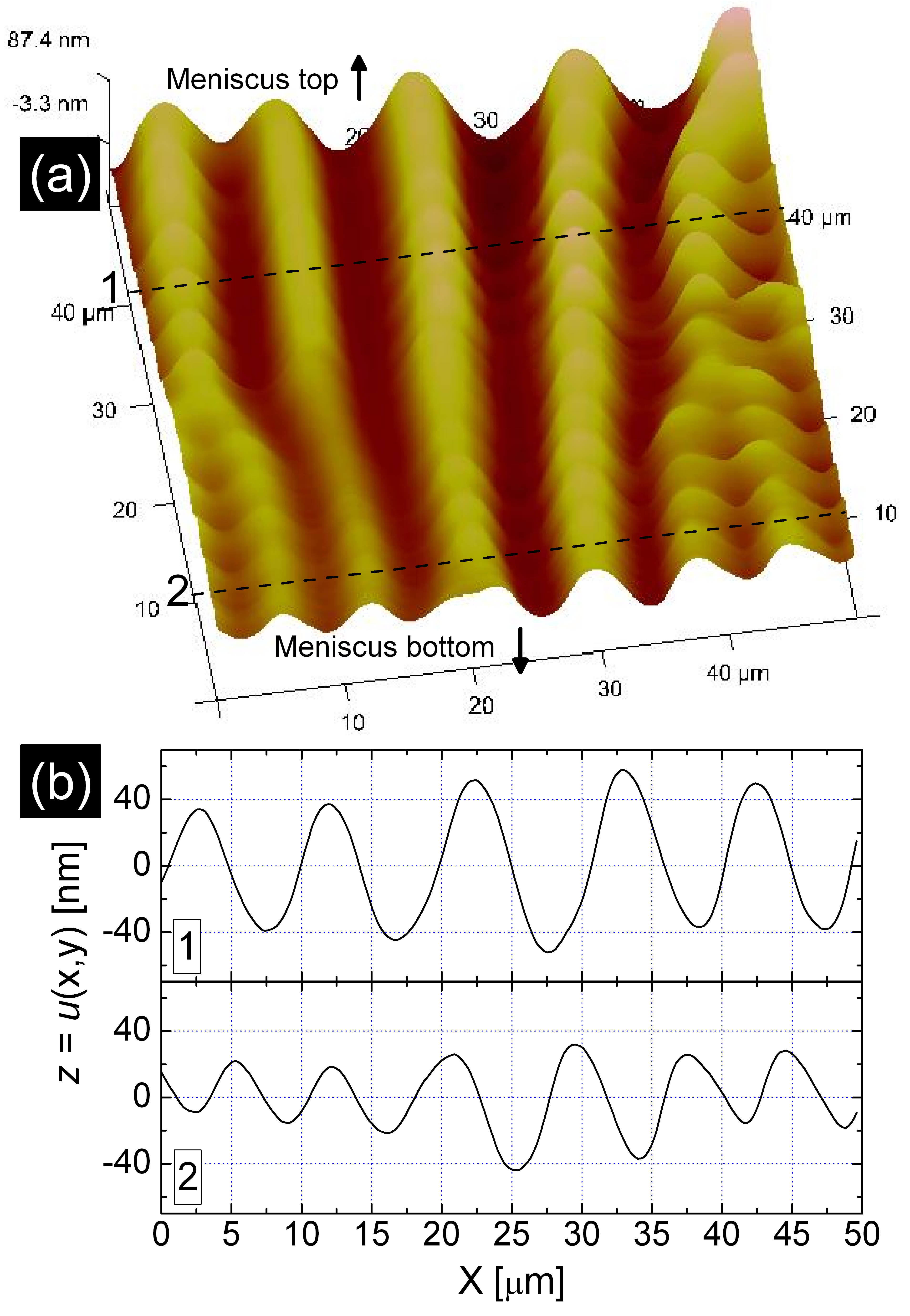}
  \end{center}
  \caption{(a) AFM image of the meniscus of a smC$^\ast$ film in the
  stripes region (SCE-12 sample, $T=20^\circ$C). The undulated
  profile clearly confirms PSI measurements. (b) $z$-profiles at
  locations marked 1 and 2 in (a).}\label{4}
\end{figure}
\end{center}

To confirm the results obtained through optical studies, we used AFM (Nanoscope Icon, Veeco, CA) to probe the
meniscus structure. To our knowledge, we report here on the first AFM experiments performed on FSF. Those very
delicate measurements were performed in air in the smC$^\ast$ phase with the microscope operated in both tapping
and peak force modes using two different tips \cite{afm}. Figure 4 exhibits the AFM results in the stripes
region. The obtained image [Fig. 4(a)] and profiles [Fig. 4(b)] undoubtedly confirm the undulation of the
smectic-air interface. Both the undulation amplitude ($h^{\mathrm{AFM}}\simeq50-100\,$nm) and wavelength
($\lambda^{\mathrm{AFM}}\simeq8-10\,\mu$m) decrease from the thick to the thin areas and are quite consistent
with those inferred from PSI. Furthermore, we have checked that the stripe period deduced from the AFM profiles
is equal to that derived from the \textit{in situ} optical image recorded just before scanning the free surface.
Note that the secondary undulations along the stripes in the AFM image do not represent actual interface
undulations but originate from scanning artefacts.

A possible mechanism that accounts for wrinkles in smC$^\ast$ menisci may be derived from the one first proposed
by Johnson and Saupe \cite{johnson77}. Assuming a constant layer number, smA layers tend to contract upon
cooling into the smC$^\ast$ phase because of an increase of the tilt angle. This contraction may not freely take
place in the meniscus because its boundaries are rigidly pinned to the solid substrate \cite{chevron}.
Therefore, a mechanical stress results and it is as if a negative pressure, or dilation, were applied to the
meniscus at the phase transition. And it is well-known that dilation-induced strain in smectics can be relaxed
by layer undulations in cases of strong anchoring \cite{johnson77,delaye73,clark73,rosenblatt77,oswpier}. In our
experiments, layers undulate at free surfaces and in the bulk of the film as well. These instabilities may very
well explain the existence of director splay domains without invoking surface-induced spontaneous polarization,
as proposed earlier \cite{meyer73,harth09}.

We can account for interfacial undulations by adding a surface tension term in the original model of Clark and
Meyer \cite{clark73}. Close to $T_C$, we may assume a very small tilt angle \cite{tiltac} and therefore use the
elastic free energy of a smA liquid crystal, as a first approximation. In terms of the layer displacement
$u(x,z)$, in the $z$-direction, and assuming small interfacial slopes, the simplest free energy density $f$
takes the form (in two dimensions)
\begin{eqnarray}
  f &=& \frac{\bar{B}}{2d\lambda}\int_{-d/2}^{d/2}dz\int_0^\lambda dx\left\{\left[\frac{\partial u}{\partial z}-
    \frac{1}{2}\left(\frac{\partial u}{\partial x}\right)^2\right]^2
    \right.\nonumber\\* & & \left.
    +\Lambda^2\left(\frac{\partial^2u}{\partial
    x^2}\right)^2\right\}+\frac{\sigma}{d\lambda}\int_0^\lambda
    dx\,h_x^2\,,
\end{eqnarray}
where $d$ is the film thickness, $q_x=2\pi/\lambda$ is the wave
vector of the instability (in the $x$-direction),
$\Lambda=\sqrt{K_u/\bar{B}}$ and $h=u|_{z=d/2,-d/2}$. The first
integral in $f$ is the usual elastic contribution describing changes
in layer thickness (described by the compression modulus $\bar{B}$)
and splay curvature (quantified by the splay elastic constant
$K_u$). The second integral is the additional surface tension term
($\sigma$ is the smectic-air interfacial tension) which will
quantify the surface energy cost associated with interfacial
deformations. At the instability threshold, we may look for a
solution $u$ of the form $u(x,z)=\gamma_0z+g(z)\cos q_xx\,$, where
$\gamma_0=\delta/d$ is the strain induced to the layers due to the
change $d\to d+\delta\,$. Minimization of $f$ with the above form of
$u$ leads to an equation for $g$ from which we get
$g(z)=g_0\cos^{-1}(q_zd/2)\cos q_zz\,$, where the symmetry
$g(z)=g(-z)$ and the condition $g(z)|_{z=d/2,-d/2}=g_0$ were used.
$q_z=q_x\sqrt{\gamma_0-\Lambda^2q_x^2}\,$. Here, $q_z$ is not a mere
constant, as in \cite{clark73}, because the layers are free to
undulate at free surfaces. Injecting the final field $u$ in $f$, the
condition for instability, $\partial f/\partial(g_0^2)\leq0\,$, is
met for $\gamma_0$ exceeding a threshold
\begin{equation}
    \gamma_0=\Lambda^2q_x^2+\frac{q_zd}{\alpha}\cot\left(\frac{q_zd}{2}\right)\,,
\end{equation}
with $\alpha$ a dimensionless parameter given by $\alpha=d\bar{B}/\sigma\,$. Because $q_z$ is a function of
$q_x$ and $\gamma_0$, the critical wavevector $q_c$ and critical strain $\gamma_{0c}$ cannot be easily
determined analytically but can be found numerically by minimizing $\gamma_0$. Taking $K_u=10^{-11}\,$N and
$\bar{B}=10^7\,\mathrm{J/m}^3$ gives $\Lambda=10^{-9}\,$m and we estimate $\sigma\simeq
2.10^{-2}\,\mathrm{J/m}^2\,$ \cite{picano00,veum06}. For a given $d$, the value of $\gamma_0$ is adjusted till
Eq. (2) is satisfied for an appropriate $q_x$ range. For $d=10\,\mu$m, for example, this calculation predicts
$\lambda_c=0.85\,\mu$m and $\gamma_{0c}=3.9\times10^{-4}\,$ at threshold, corresponding to a critical
displacement $\delta_c=3.9\,$nm which is independent of film thickness, as in \cite{clark73}. The computed value
of $\lambda_c$ is of the same order of magnitude than the experimental one derived at threshold,
$\lambda_c^{\mathrm{exp}}\simeq 1.5-2\,\mu$m. Furthermore, our calculations show that $\lambda_c$ increases with
$d$ (more precisely, $\lambda_c\propto\sqrt{d}$, results not shown), which is also consistent with the
experimentally observed trend (Fig. 2). The dilation is greater in thicker parts of the meniscus because more
layers are involved. Therefore, a dilation gradient occurs in the meniscus, suggesting that the 2D square
lattice develops only when a high enough dilative strain has been reached. It is then very tempting to think
that this lattice corresponds to the network of parabolic focal conics (PFC) defects mentioned in earlier
studies \cite{picano00,oswpier,rosenblatt77,delrieu74}. Indeed, PFC's nucleate in (highly) dilated smA samples
to (partly) relax the stress. This is precisely what seems to occur here in the meniscus.

Neither the stripes nor the 2D pattern appear in the flat film
because no layer dilation occurs there at the phase transition, at
least within our experimental conditions (i.e. for $d\leq 1\,\mu$m).
Indeed, if layers contract sufficiently below $T_C$, it is
well-known that smectic layers can sustain compression without
undulating \cite{oswpier} and we have checked this fact within our
simple model ($\gamma_0<0$). Note, however, that Gorecka \textit{et
al.} \cite{gorecka95} observed instabilities in free standing
smectic films but only for thick enough films ($d\geq5\,\mu$m).

Finally, in contrast to instabilities in sandwiched cells, the
structures in smC$^\ast$ menisci are stable for weeks. We may expect
the numerous dislocations in the meniscus to rearrange locally to
relax the stress. Such plastic relaxation mechanisms are indeed very
likely \cite{oswpier} but fail to explain the long term stability of
meniscus patterns.

To conclude, our results provide clear-cut evidences that the meniscus structure in smC$^\ast$ (smC) samples is
more complex than that presumed earlier as it is characterized by undulations of the smectic layers. Additional
experiments are currently in progress to investigate the influence of film thickness and the behavior of
colloidal inclusions. On the theory side, a more sophisticated model involving thickness gradients and the
coupling of $\mathbf{c}$-director distortions to layer undulations should be worked out.

The authors cordially acknowledge L. Lej\v{c}ek and B. Pouligny for very helpful discussions and thank J. P.
Salvetat for suggesting the AFM experiments.


\begin{thebibliography}{99}
\bibitem{2dnet} Note that this 2D pattern is actually visible in figs. 4 and 6 of ref. \cite{harth09} but is not
  mentionned nor discussed in the text.
\bibitem{afm} In the tapping mode, phase images were obtained using a silicon tip (stiffness $k\simeq10\,$N/m,
  tip radius $<10\,$nm, $\nu=150\,$kHz) whereas $\mathrm{Si}_3\mathrm{N}_4$ tips
  (stiffness $k\simeq0.4\,$N/m, tip radius $<10\,$nm) were used for imaging in the peak force mode.
\bibitem{tiltac} This is true only if the smA-smC phase transition is of second order.
\end{thebibliography}
\end{document}